\documentclass[twocolumn,showpacs,showkeys,byrevtex,pra]{revtex4}%
\usepackage{amsfonts}
\usepackage{amsmath}
\usepackage{amssymb}
\usepackage{graphicx,color}%
\providecommand{\U}[1]{\protect\rule{.1in}{.1in}}
\begin{document}
\preprint{ }
\title{Information trade-offs for optical quantum communication}
\author{Mark M. Wilde,$^1$ Patrick Hayden,$^1$ and Saikat Guha$^2$}
\affiliation{$^1$\textit{School of Computer Science, McGill University, Montreal, Qu\'{e}bec,
Canada H3A 2A7},\\
$^2$\textit{Disruptive Information Processing Technologies Group, Raytheon BBN
Technologies, Cambridge, Massachusetts, USA 02138}}
\keywords{trade-off coding, quantum Shannon theory, bosonic channels, entanglement,
secret key}
\pacs{03.67.Hk, 03.67.Pp, 04.62.+v}

\begin{abstract}
Recent work has precisely characterized the achievable trade-offs between three key information processing tasks---classical communication (generation or consumption), quantum communication (generation or consumption), and shared entanglement (distribution or consumption), measured in bits, qubits, and ebits per channel use, respectively. Slices and corner points of this three-dimensional region reduce to well-known protocols for quantum channels. A trade-off coding technique can attain any point in the region and can outperform time-sharing between the best-known protocols for accomplishing each information processing task by itself. Previously, the benefits of trade-off coding that had been found were too small to be of practical value (viz., for the dephasing and the universal cloning machine channels). In this letter, we demonstrate that the associated performance gains are in fact remarkably high for several physically relevant bosonic channels that model free-space / fiber-optic links, thermal-noise channels, and amplifiers. We show that significant performance gains from trade-off coding also apply when trading photon-number resources between transmitting public and private classical information simultaneously over secret-key-assisted bosonic channels.
\end{abstract}
\date{\today}
\startpage{1}
\endpage{10}
\maketitle

Shannon's classical information theory found the capacity of a classical channel, which quantifies the channel's ability to transmit information~\cite{bell1948shannon}. The capacity serves as a benchmark against which communication engineers can test the performance of any practical scheme. Despite the success of Shannon's theory, it fails to identify the true capacity for physical channels such as free-space or fiber-optic links because the quantum physical properties of the optical-frequency EM waves---the carriers of information---must be accounted for within a full quantum framework in order to assess the ultimate limits on reliable communication~\cite{S09}. A major revision of Shannon's information theory, dubbed \textit{quantum Shannon theory}, has emerged in recent years in an attempt to determine the ultimate physical limits on communications~\cite{BS04}. This theory has provided a successful quantum theory of information in many special cases~\cite{HP01,K03,DS03,GGLMSY04,WPG07}, but recent developments have indicated that there is much more to understand regarding the nature of information transmission over quantum channels~\cite{science2008smith,H09}.
%The quantum nature of light must be correctly accounted for in order to assess the ultimate performance limits of information processing systems that use optical-frequency EM waves. {\em Quantum Shannon theory}~\cite{BS04}---a major revision of Shannon's classical information theory~\cite{bell1948shannon}---aims to determine the ultimate limits on communications that leverage quantum resources~\cite{BS04}.
%%Despite the success of this theory in many special cases~\cite{HP01,K03,DS03,GGLMSY04,WPG07}, several nuances of the nature of information transmission over quantum channels still elude us~\cite{science2008smith,H09}.

Quantum channels support a richer variety of information processing tasks than do classical channels. A sender can transmit classical information, such as \textquotedblleft on\textquotedblright\ or \textquotedblleft off,\textquotedblright~\cite{HJSWW96,SW97,ieee1998holevo}, or she can transmit quantum information, such as the quantum state of a photon~\cite{L97,capacity2002shor,ieee2005dev}. Additionally, if the sender and receiver have prior shared entanglement, this resource can boost the rate of information transmission~\cite{BSST01}, generalizing the superdense coding effect~\cite{BW92}. The sender might also want to transmit classical and quantum information simultaneously to the receiver~\cite{DS03}, or even limit the amount of entanglement consumed in the entanglement-assisted transmission of classical and/or quantum information~\cite{S04}. The sender and receiver could further specify whether they would like the classical information to be public or private~\cite{WH10a}.

In a ``trade-off" communication problem, such as that of simultaneous classical-quantum communication, a naive strategy of time-sharing would have the sender and receiver use a classical communication protocol for some fraction of the time (say, the best Holevo-Schumacher-Westmoreland (HSW) classical code and a joint-detection receiver on long codeword blocks~\cite{HJSWW96, SW97, ieee1998holevo}), while operating with a quantum communication protocol for the other fraction of the time (say, the best Lloyd-Shor-Devetak (LSD) quantum code and a joint-detection receiver on quantum codewords~\cite{L97,capacity2002shor,ieee2005dev}). Trade-off coding is a more complex strategy, in which---simply stated---the sender encodes classical information into the many different ways of permuting quantum codes. Its performance can beat that of time-sharing for certain channels such as dephasing and universal cloning machine channels, for which it is even provably optimal~\cite{DS03,WH10,BHTW10,JBW11}. The original~\cite{BS04,DS03,S04} and subsequent developments~\cite{HW08ITIT,BHTW10} on trade-off coding have greatly enhanced our understanding of communication over quantum channels. However, the pay-off of trade-off coding for the channels studied previously was too small to be worthwhile in a practical setting, given the increased encoding and decoding complexity over time-sharing.

In this letter, we show that trade-off coding yields remarkable gains over time-sharing for  the single-mode lossy bosonic channel, which can model free-space optical communication. These single-mode results are sufficient to construct trade-off capacity results for any physical optical communication channel that modulates multiple degrees of freedom of the photon, such as spatial and polarization modes of light. Our results also apply more generally to thermal-noise and amplifying bosonic channels, which can model systems as diverse as superconducting transmission lines in the microwave range~\cite{Wetal04} or hybrid quantum memories that store both classical and quantum information in the collective degrees of freedom of atomic ensembles~\cite{Cetal05}. We determine an achievable rate region for the lossy bosonic channel using a transmitter that modulates the two-mode squeezed-vacuum---an entangled light state that can be generated using parametric downconversion---and prove that this rate region is optimal assuming a long-standing minimum output entropy conjecture is true~\cite{GGLMS04,GHLM10,GSE07,G08}. Even if the conjecture is not true, our achievable trade-off region beats time-sharing between the best-known quantum communication protocols by huge margins. The same holds for the thermal and amplifying channels.

\begin{figure}
\centering
\includegraphics[width=\columnwidth]{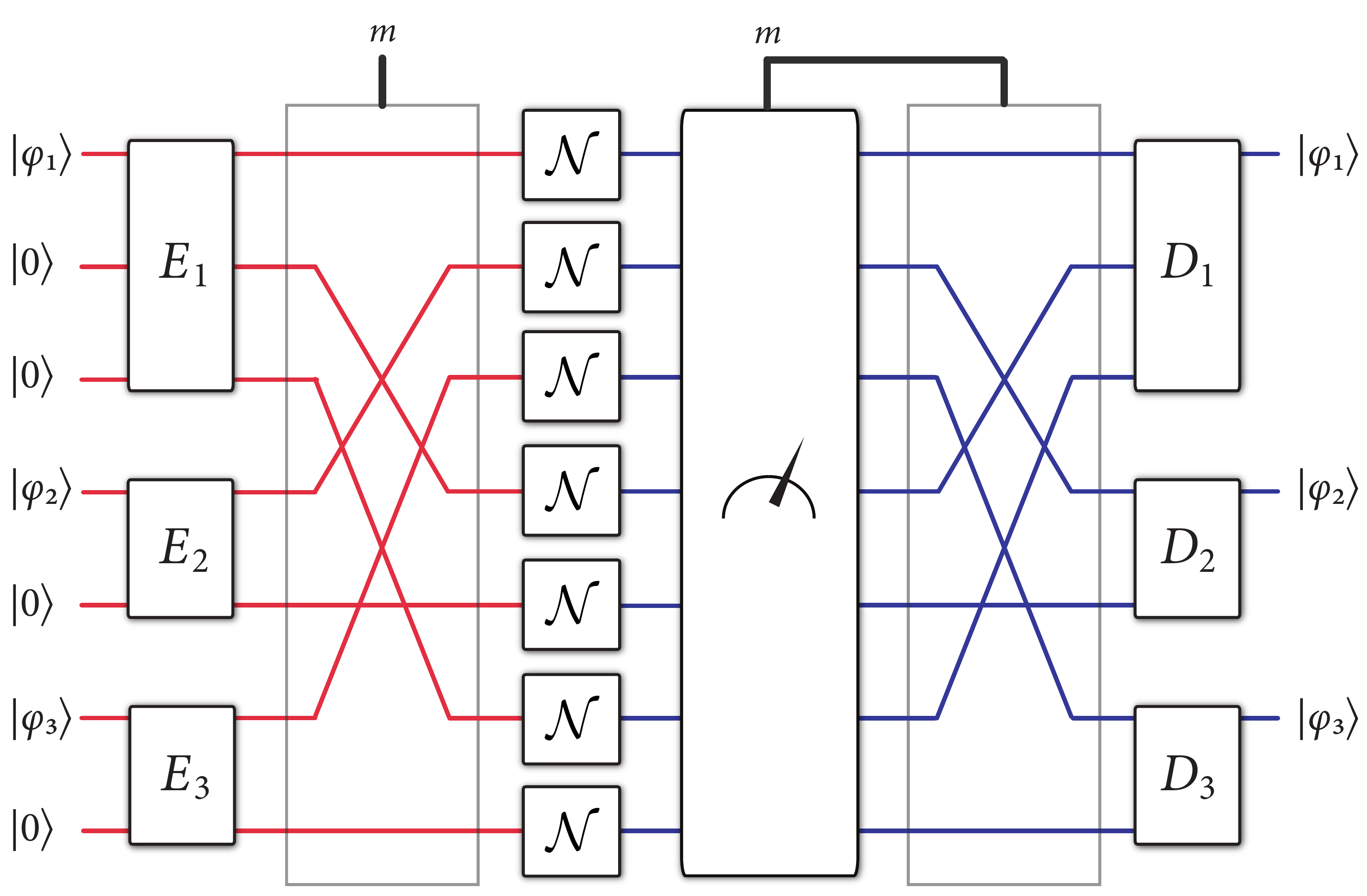}
\caption{A sketch of the trade-off coding protocol for communication of classical and quantum information
without any entanglement assistance (this case is the Devetak-Shor protocol \cite{DS03}). The sender begins by encoding qubits $\vert \varphi_1 \rangle$, $\vert \varphi_2 \rangle$, and $\vert \varphi_3 \rangle$ into different quantum error-correcting codes, each constructed from a particular state $\vert \phi_x \rangle$ and the channel $\mathcal{N}$. To encode classical information, the sender permutes the quantum systems emerging from the outputs of the encoders according to some classical message $m$. The sender then transmits these systems through many independent uses of the noisy channel $\mathcal{N}$. The receiver obtains the outputs of channels, and performs an HSW measurement to determine the classical message $m$. If the success probability of the measurement is asymptotically close to one, then it causes an asymptotically negligible disturbance to the state on which it acts. The receiver then knows the permutation, unpermutes the quantum systems, and exploits the decoders of the quantum error-correcting codes to decode the qubits $\vert \varphi_1 \rangle$, $\vert \varphi_2 \rangle$, and $\vert \varphi_3 \rangle$. The Wilde-Hsieh protocol \cite{HW10} extends this idea by permuting entanglement-assisted quantum codes in a similar way.}
\label{fig:Devetak-Shor-prot}
\end{figure}

%\begin{figure}
%\begin{center}
%\includegraphics[width=0.9\columnwidth]{EACQ-protocol.pdf}
%\caption{{\bf The anatomy of a trade-off code.} Alice and Bob are spatially separated
%(depicted with the dashed lines). Alice would like to send both classical and quantum information to Bob with the help of shared entanglement (depicted at the lower left as arising from some third-party source). Alice feeds a qubit stream
%$\left\vert \uparrow\right\rangle \left\vert \searrow\right\rangle \left\vert
%\downarrow\right\rangle \left\vert \nearrow\right\rangle \cdots$, a classical bit stream $0101100\cdots$, and her half of the shared entanglement through an encoder $\mathcal{E}$. She transmits the encoded data over many uses of a noisy channel. Bob feeds the received codeword along with his half of the shared entanglement
%into a decoder $\mathcal{D}$ that can decode both the classical and quantum data.}
%\label{fig:trade-off-anatomy}
%\end{center}
%\end{figure}

\textit{Trading quantum and classical resources}---Our first result concerns
the transmission of classical and quantum information over a single-mode
lossy bosonic channel of input-output power transmissivity $\eta \in (0, 1]$, with a constraint on the mean photon number $N_S$ per mode at the transmitter. Recall that this channel has the following input-output Heisenberg-picture specification: $\hat{b} = \sqrt{\eta} \hat{a} + \sqrt{1-\eta} \hat{e} $, where $\hat{a}$, $\hat{b}$, and $\hat{e}$ are the respective bosonic annihilation operators representing the sender's input mode, the receiver's output mode, and an environmental input in the vacuum state. Transmission over this channel
can also be used to generate shared entanglement between the sender and the receiver or this resource might assist transmission if they share it beforehand. We let $C$ be the rate of classical communication, $Q$ be the rate of quantum communication, and $E$ be the rate of entanglement generation (or consumption). If the rate of a resource is positive, then the interpretation is that the protocol generates that resource. Otherwise, the protocol consumes that resource.
%Figure~\ref{fig:trade-off-anatomy} depicts a particular scenario in which Alice would like to communicate both classical and quantum data to Bob with the help of shared entanglement ($C>0$, $Q>0$, $E<0$).

Hsieh and Wilde described a general-purpose protocol
for entanglement-assisted communication of classical and quantum
information over many independent uses of {\it any} noisy quantum channel~\cite{HW08ITIT} and subsequently found the full $(C,Q,E)$ {\em triple trade-off region}~\cite{HW10,WH10}.
The Hsieh-Wilde protocol is constructed from a particular ensemble $\left\{  p_{X}\left(  x\right)  ,\rho_{x}\right\}  $
and the channel $\mathcal{N}$. Let $\vert \phi_x \rangle$ denote a purification of $\rho_x$, let $\rho\equiv\sum_{x}p_{X}\left(  x\right) \rho_{x}$
be the average density operator of the ensemble, 
  let $\mathcal{N}^{c}$ be the channel complementary to $\mathcal{N}$ \footnote{Every quantum channel $\mathcal{N}$ has an isometric extension to a larger Hilbert space \cite{DS03}. One could imagine this isometric extension arising operationally from a unitary interaction between the sender's input system and an environment initialized in some pure state. After the unitary acts on the joint state, one obtains the original noisy channel to the receiver by tracing over the environment system, and one obtains the complementary channel to the environment by tracing over the receiver's system.}, and $H(\sigma)\equiv -\operatorname{Tr} ( \sigma \log_2 \sigma )$ the von Neumann entropy. Then
the Hsieh-Wilde protocol generates $H\left(  \mathcal{N}\left(  \rho\right)  \right)
-\sum_{x}p_{X}\left(  x\right)  H\left(  \rho_{x}\right)$ bits per channel use and
$ \sum_{x}p_{X}\left(  x\right) [ H\left(  \rho_x\right)
+    H\left(  \mathcal{N}(\rho_{x})\right)  -H\left(  \mathcal{N}^{c}\left(  \rho_{x}\right)  \right)
]/2$ qubits per channel use by consuming 
$ \sum_{x}p_{X}\left(  x\right) [ H\left(  \rho_x\right)
+ H\left(  \mathcal{N}^{c}\left(  \rho_{x}\right)  \right) -   H\left(  \mathcal{N}(\rho_{x})\right)
]/2$ ebits per channel use \cite{HW08ITIT}. This protocol is a {\it trade-off coding protocol}, in the sense that
encoded classical and quantum data can be fed into the same channel input, rather than into
separate channel inputs, as is the
case in a time-sharing protocol that allocates a portion of the channel uses solely for classical data tranmission and the other portion solely for quantum data transmission.
Figure~\ref{fig:Devetak-Shor-prot} depicts the operation of this protocol in the case where there is no entanglement assistance.
Combining the Hsieh-Wilde protocol with teleportation, super-dense coding,
and entanglement distribution (while keeping track of net rates) gives the following achievable rate region \cite{WH10}:
\begin{gather}
C+2Q    \leq H\left(  \mathcal{N}\left(  \rho\right)  \right)
+\sum_{x}p\left(  x\right)  \left[  H\left(  \rho
_{x}\right)  -H\left(  \mathcal{N}^{c}\left(  \rho_{x}\right)  \right)
\right]  ,\nonumber\\
Q+E    \leq\sum_{x}p\left(  x\right)  \left[  H\left(  \mathcal{N}\left(
\rho_{x}\right)  \right)  -H\left(  \mathcal{N}^{c}\left(  \rho_{x}\right)
\right)  \right]  ,\nonumber\\
C+Q+E    \leq H\left(  \mathcal{N}\left(  \rho\right)  \right)  -\sum
_{x}p\left(  x\right)  H\left(  \mathcal{N}^{c}\left(  \rho_{x}\right)
\right)  . \label{eq:CQE-region}%
\end{gather}
Hsieh and Wilde also proved a multi-letter converse, so that the above region's regularization is optimal \cite{HW10,WH10}.

For the lossy bosonic channel, the Hsieh-Wilde protocol and rate region translate to the following. The protocol is constructed from
an ensemble of Gaussian-distributed phase-space displacements of two-mode squeezed vacuum (TMSV) states:
$
\{  p_{\overline{\lambda}N_{S}}\left(  \alpha\right)  ,D^{A^{\prime}%
}\left(  \alpha\right)  |\psi_{\text{TMS}}\rangle^{AA^{\prime}}\}  
%\label{eq:bosonic-ensemble}%
$, where $A'$ is a system sent into the channel input and $A$ is a system that purifies $A'$.
(In the above, the distribution
$p_{\overline{\lambda}N_{S}}\!\left(  \alpha\right)$ replaces $p_X(x)$, and the state $D^{A^{\prime} 
}\!\!\left(  \alpha\right)  |\psi_{\text{TMS}}\rangle^{AA^{\prime}}$ replaces $\vert \phi_x \rangle$.)
The distribution
$
p_{\overline{\lambda}N_{S}}\left(  \alpha\right)  \equiv\frac{1}{\pi
\overline{\lambda}N_{S}}\exp\{  -\left\vert \alpha\right\vert
^{2}/\overline{\lambda}N_{S}\}
$
is an isotropic Gaussian distribution with variance $\overline{\lambda}N_{S}$,
where $\overline{\lambda}\equiv1-\lambda$ and $\lambda\in\left[ 0,1\right]  $ is
a photon-number-sharing parameter.
%indicating how many photons to dedicate to
%the quantum part of the code, while $\overline{\lambda}$ indicates how many
%photons to dedicate to the classical part.
The state $|\psi_{\text{TMS}}\rangle^{AA^{\prime}}$ is a two-mode
squeezed vacuum~\cite{GK04,BVL05}:
\begin{equation}
\!|\psi_{\text{TMS}}\rangle^{AA^{\prime}}\equiv\sum_{n=0}^{\infty}\sqrt
{\left[  \lambda N_{S}\right]  ^{n}/\left[  \lambda N_{S}+1\right]
^{n+1}}\left\vert n,n\right\rangle ^{AA^{\prime}}.
\label{eq:two-mode-squeezed}%
\end{equation}
Evaluating the entropies in (\ref{eq:CQE-region}) for this ensemble and the lossy bosonic channel gives the following achievable rate region: 
\begin{align}
C+2Q  &  \leq g\left(  \lambda N_{S}\right)  +g\left(  \eta N_{S}\right)
-g\left(  \left(  1-\eta\right)  \lambda N_{S}\right)  ,\nonumber\\
Q+E  &  \leq g\left(  \eta\lambda N_{S}\right)  -g\left(  \left(
1-\eta\right)  \lambda N_{S}\right)  ,\nonumber\\
C+Q+E  &  \leq g\left(  \eta N_{S}\right)  -g\left(  \left(  1-\eta\right)
\lambda N_{S}\right)  , \label{eq:CQE-lossy-bosonic}%
\end{align}
where $N_{S}$ is the input mean photon number per mode (per channel use), and $g\left(N\right)$ is the entropy of a single-mode thermal state with mean photon number $N$: $g\left(N\right)  \equiv\left(  N+1\right)  \log_2 \left(  N+1\right)  -N\log_2 N$.
The photon-number sharing parameter~$\lambda$ is the fraction of photons the code dedicates to quantum resources
%(such as quantum communication or entanglement-assisted classical transmission)
as compared to classical resources.
%(such as unassisted classical communication).
This allocation, however, is done within a single channel use (as a power-sharing strategy), whereas in a time-sharing strategy, each channel use is dedicated to only one task at a time. The above result extends to other important bosonic channels such as the thermal-noise and amplifier channels~\cite{WHG11}.

\begin{figure}
\centering
\includegraphics[width=\columnwidth]{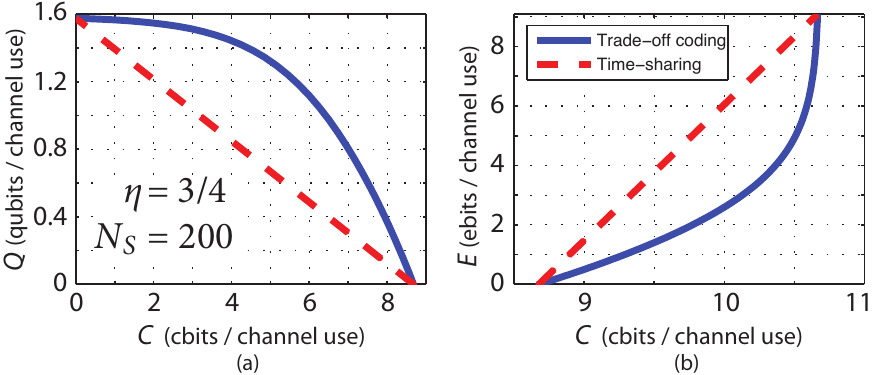}
\caption{(a) {\bf The $\mathbf{(C,Q)}$ trade-off.} A lossy bosonic channel with transmissivity $\eta = 3/4$ can reliably transmit a maximum of $\log_{2}\left(  3/4\right)  -\log_{2} \left(  1/4\right)  \approx 1.58$ qubits per channel use~\cite{WPG07}, and $N_S=200$~photons per mode at the channel input is sufficient to nearly achieve this quantum capacity. A trade-off coding strategy that lowers the quantum data rate to about $1.4$~qubits per use while
retaining the same mean photon budget, allows the transmission of an additional $4.5$~classical bits per channel use, while time-sharing would only allow for an additional $1$~classical bit per channel use with this photon budget. (b) {\bf The $\mathbf{(C,E)}$ trade-off.} The sender and the receiver share entanglement, and the sender would like to transmit classical information while minimizing the consumption of entanglement. With a
mean photon budget of $N_S=200$~photons per channel use, the sender can reliably
transmit a maximum of about $10.7$~classical bits per channel use while
consuming entanglement at a rate of about $9.1$~entangled bits
per channel use~\cite{BSST01,HW01,GLMS03}. With trade-off coding, the sender can
significantly reduce the entanglement consumption rate to about $5$~entangled
bits per channel use while still transmitting about $10.5$~classical bits per
channel use, i.e., only a $0.08$~dB decrease in the rate of classical communication
for a $2.6$~dB decrease in the entanglement consumption rate.}
\label{fig:bosonic-trade-offs}
\end{figure}

%The RHS of the inequalities in \eqref{eq:CQE-lossy-bosonic} can be divided by $\ln 2$, such that the units for $C$, $Q$ and $E$ become bits, qubits, and entangled bits (ebits) per channel use, respectively.
Note that $Q=0$ for $\eta < 1/2$, thereby making the $(C,Q)$ trade-off region trivial for $\eta < 1/2$. However, for the $(C,E)$ trade-off, trade-off coding with the Hsieh-Wilde protocol outperforms time-sharing for all values of $\eta$.

If $\eta \geq 1/2$ and the minimum output entropy conjecture is true, then the rate region defined by (\ref{eq:CQE-lossy-bosonic}) is the actual capacity region~\cite{WHG11}.
Our proof of optimality is similar to the optimality proof for the
bosonic broadcast channel \cite{GSE07}, a setting in which a sender, at one input port of a beamsplitter transmits classical data to two receivers at the two output ports. For the noiseless broadcast channel, the second beamsplitter input is in the vacuum state. In the broadcast setting, there is always one receiver whose output is less noisy than the other's (whichever receiver has the output for which $\eta \geq 1/2$).
The techniques for proving optimality of rates to the less noisy receiver readily apply when analyzing
our setting \cite{WHG11}, but we require $\eta \geq 1/2$
in order to apply them because there is only one receiver in our setting.

%\begin{figure}[h]
%\begin{center}
%\includegraphics[
%%natheight=4.373400in,
%%natwidth=7.052600in,
%%height=2.0816in,
%width=\columnwidth
%]%
%{CQE-eta-3-4-NS-100.pdf}%
%\caption{The full triple trade-off region for the lossy bosonic channel with
%transmissivity $\eta=3/4$ and mean input photon number $N_{S}=100$.
%The arrows indicate directions in which the boundary extends infinitely.}%
%\label{fig:CQE-bosonic}%
%\end{center}
%\end{figure}
%Figure~\ref{fig:CQE-bosonic} depicts an example evaluation of the full triple trade-off (CQE) region for the lossy bosonic channel with $\eta = 0.75$ and $N_S = 100$ photons.

Figure~\ref{fig:bosonic-trade-offs} depicts two important special cases of the region in~(\ref{eq:CQE-lossy-bosonic}):\ (a) the trade-off between classical and quantum communication without entanglement assistance and (b) the trade-off between entanglement-assisted and unassisted classical communication. The figure indicates the remarkable improvement over time-sharing that trade-off coding achieves for the lossy bosonic channel. If $N_S$ is high enough to achieve close to the maximum quantum capacity $\log_2(\eta) - \log_2(1-\eta)$, then the achievable rates in \eqref{eq:CQE-lossy-bosonic} are much better than those achievable by time-sharing between its quantum capacity~\cite{WPG07}, its classical capacity~\cite{GGLMSY04}, and its entanglement-assisted classical and quantum capacities~\cite{BSST01,HW01,GLMS03}. 

\textit{A rule of thumb for trade-off coding}---The quantum capacity of a lossy bosonic channel with transmissivity $\eta$ and mean photon number per mode $N_S$ is given by $Q(\eta,N_S) = \max\left[0, g\left(  \eta N_{S}\right)  -g\left(  \left(  1-\eta\right)  N_{S}\right) \right]$ \cite{HW01,WPG07,GSE08}. Note that $Q(\eta,N_S) = 0$, $\forall \eta \le 1/2$, and $\lim_{N_S \to \infty}Q(\eta, N_S) = \log_2\left(  \eta\right)  -\log_2\left(1-\eta\right) \equiv Q_{\max}(\eta)$. In the context of trade-off coding, the achievable rate becomes $Q(\eta,{\lambda}N_S)$, where $\lambda$ is the fraction of photons dedicated to quantum resources. A Taylor series expansion yields $Q(\eta,{\lambda}N_S) \ge Q_{\max}(\eta) -\left[  \eta\left(  1-\eta\right)\lambda N_{S}\ln2\right]  ^{-1}$ when $N_{S}$ is sufficiently high \cite{WHG11}. Thus, in order to reach the quantum capacity to within
$\epsilon$ bits, a trade-off code should dedicate no more than a fraction
$\lambda^*=1/\left[  \eta\left(  1-\eta\right)  \epsilon N_{S}\ln2\right]  $ to
the quantum part of the code. \textit{If the trade-off code dedicates a higher
fraction of the photons than $\lambda^*$ to quantum resources, then it is effectively
wasting photons which could instead be used to get a significant amount of
classical communication \textquotedblleft for free}.\textquotedblright\ As $N_S$ increases, the fraction of available photons needed
for the quantum rate to saturate at $Q_{\max}(\eta)$ becomes smaller and smaller. So,
if a trade-off code abides by the above rule of thumb, it will nearly saturate
$Q_{\max}(\eta)$, while achieving a high classical data rate as well---something that is {\em not} possible by merely time sharing between classical (HSW) and quantum (LSD) communication. A similar rule of thumb applies for entanglement-assisted classical communication, i.e., it is not necessary to dedicate a large
fraction of the photons to shared entanglement when the photon budget
increases~\cite{WHG11}.% (we refer the reader to \cite{WHG11} for more details).

%Recall that the maximum classical communication rate over the lossy
%bosonic channel with entanglement assistance is equal to $g\left(
%N_{S}\right)  +g\left(  \eta N_{S}\right)  -g\left(  \left(  1-\eta\right)
%N_{S}\right)  $~\cite{HW01,GLMS03}, while from the first inequality
%in~(\ref{eq:CQE-lossy-bosonic}), observe that a limited-entanglement trade-off
%code achieves a rate of $g\left(  \lambda N_{S}\right)  +g\left(  \eta
%N_{S}\right)  -g\left(  \left(  1-\eta\right)  \lambda N_{S}\right)  $. Thus,
%a natural question is to determine the fraction of photons needed to dedicate
%to shared entanglement in order to come close to achieving the full
%entanglement-assisted classical capacity (with maximal entanglement
%consumption). It turns out that we can apply a similar rule of thumb because
%the quantity $5/\left[  6\lambda N_{S}\left(  1-\eta\right)  \right]  $ serves
%as an upper bound on the difference of the entanglement-assisted capacity and
%the limited entanglement data rate when $N_S$ is sufficiently high~\cite{WHG11}. Thus, in order to be within
%$\epsilon$ bits of the entanglement-assisted classical capacity, it suffices
%to choose the photon-number sharing parameter $\lambda=5/\left[  6\epsilon
%N_{S}\left(  1-\eta\right)  \ln2\right]  $~\cite{WHG11}. Since the rate of entanglement
%consumption for a limited entanglement code is $g\left(  \lambda N_{S}\right)
%$, this rule of thumb can give a significant savings on entanglement
%consumption in order to have the benefit of the superdense coding effect in
%the high photon-number regime.

\textit{Trading public and private classical resources}---Analogous
trade-off coding results hold for another notable setting,
where a sender would like to transmit both public and private
classical information to a receiver over a bosonic channel (perhaps even with
the assistance of a secret key). These results constitute a relevant benchmark
for satellite-to-satellite (far-field free-space) links, which might be used for both public
communication and quantum key distribution~\cite{SBCDLP09}. We let $R$~denote
the rate of public communication, $P$~the rate of private communication, and
$S$~the rate of secret key generation/consumption. An achievable rate region for the
lossy bosonic channel with $\eta \in [0,1]$ is
\begin{align}
R+P  &  \leq g\left(  \eta N_{S}\right)  ,\nonumber\\
P+S  &  \leq g\left(  \eta\lambda N_{S}\right)  -g\left(  \left(
1-\eta\right)  \lambda N_{S}\right)  ,\nonumber\\
R+P+S  &  \leq g\left(  \eta N_{S}\right)  -g\left(  \left(  1-\eta\right)
\lambda N_{S}\right)  , \label{eq:RPS-lossy-bosonic}%
\end{align}
where $\lambda\in\left[  0,1\right]  $, a photon-number-sharing parameter, is the fraction of photons dedicated to private classical resources, and $N_{S}$ is the mean input photon number per mode. If $\eta \geq 1/2$  and the minimum-output entropy conjecture is true, then this region is the capacity region \cite{WHG11}.
We were able to prove optimality here again by appealing to the optimality results from the bosonic broadcast channel \cite{GSE08}.
For $\eta < 1/2$, the above region remains achievable. The strategy for achieving the above rate region is to combine the general-purpose Hsieh-Wilde protocol for secret-key-assisted communication of public and private classical information \cite{WH10a} and combine it with the one-time pad, secret key distribution, and private-to-public transmission (the ideas here are similar to those for the CQE trade-off). For the lossy bosonic channel, coherent-state codewords selected according to an isotropic Gaussian distribution
suffice to achieve the above region~\cite{WHG11}.

An interesting special case of the above achievable region is the trade-off
between public and private classical communication. Lemma~3 of
Ref.~\cite{WH10a} proves that the classical-quantum trade-off region is the same as the public-private trade-off whenever the channel is degradable, which applies here since
the lossy bosonic channel is degradable whenever $\eta\geq1/2$~\cite{WPG07,GSE08}.
These special cases coincide because ensembles of pure states suffice
for achieving the private classical capacity of degradable channels \cite{WH10a},
further implying in such a case that the private information is equivalent to the coherent information. Thus, Figure~\ref{fig:bosonic-trade-offs}(a) doubles as a plot of the public-private trade-off ($C \to R, Q \to P$). Furthermore, note that the trade-off between public
classical communication and secret key generation is the same as that between
public and private classical communication, respectively.

%Another setting to which our results apply is more exotic---that of relativistic optical quantum communication. If a receiver accelerates uniformly with respect to an inertial sender, then
%the sender's transmission to the receiver experiences an amplification noise
%due to the well-known Unruh effect~\cite{PhysRevD.14.870}. Prior research has
%investigated this \textquotedblleft Unruh channel\textquotedblright\ under a
%somewhat restrictive assumption that the sender transmit certain superpositions of single-photon Fock states to the receiver~\cite{BHP09,BHTW10,JBW11}. Our results demonstrate an improvement in performance for this Unruh channel under the less restrictive assumption of a fixed photon budget, with the upshot of a considerable simplification in the
%mathematics needed to understand the communication capabilities of this
%channel~\cite{WHG11}.

\textit{Discussion}---We might attempt to understand why trade-off coding
between classical and quantum communication performs so well for bosonic
channels in the high photon-number regime by making an analogy with qubit
dephasing channels. It is well known that for every $\eta < 1$, the
lossy bosonic channel has a finite quantum capacity even when an infinite number of
photons are available~\cite{HW01}. Here, we have seen that we can approach
this quantum capacity with just a small fraction of the total photon number
dedicated to the quantum part of the code. Thus, one can think loosely of the
lossy bosonic channel as being \textquotedblleft composed of\textquotedblright\ a
few channels that are good for quantum transmission while the rest are
good for classical transmission. The analogy is that we can combine many
strongly dephasing channels that are only good for classical data transmission
with just a few weakly dephasing channels that are good for quantum data
transmission in order to approximate the lossy bosonic classical-quantum
trade-off. Refs.~\cite{DS03,BHTW10,HW08ITIT} prove that the trade-off
capacity region of any dephasing channel is additive, and so the resulting
region for the combined dephasing channel is simply the Minkowski sum of those of the
individual channels. However, this understanding is only satisfying
in the very high photon-number regime, when the available number of photons is
much larger than that needed to saturate the quantum capacity.

\textit{Conclusion}---We have shown that achievable rates with trade-off coding
over bosonic channels can be significantly higher than those achievable from time-sharing between conventional quantum protocols, suggesting that quantum communication engineers should try to take advantage of these gains in a practical coding scheme. Our trade-off regions are optimal for a lossy bosonic channel that transmits on average over half of the photons input to it, assuming that the minimum-output entropy conjecture is true. This paper does not discuss specific codes and structured optical receivers to attain reliable communications at rate-triples predicted by our achievable trade-off capacity region. In future work, it would be interesting to lay out the full transmitter-coding-receiver architecture for optical trade-off coding.

\textit{Acknowledgements}---We thank J.~H.~Shapiro for reminding us
of relevant results~\cite{GSE07}. MMW acknowledges the MDEIE\ (Qu\'{e}bec) PSR-SIIRI
international collaboration grant. PH\ acknowledges the hospitality of the Stanford Institute
for Theoretical Physics as well as funding from the Canada Research Chairs
program, the Perimeter Institute, CIFAR, FQRNT's INTRIQ, NSERC, ONR
through grant N000140811249, and QuantumWorks. SG acknowledges the DARPA
Information in a Photon program, contract \#HR0011-10-C-0159.

\bibliography{Ref}
\bibliographystyle{unsrt}

\end{document}